\documentclass[11pt,twoside]{article}

\usepackage{asp2006}
\usepackage{epsf}
\usepackage{psfig}
\usepackage{lscape}
\usepackage{graphicx}

\begin{document}
\title{What can we learn on the structure and the dynamics of the solar core with $g$ modes?}   
\author{S. Mathur}   
\affil{Indian Institute of Astrophysics, Koramangala, Bangalore 560034, India}    
\author{J. Ballot}
\affil{Max-Planck-Institut f\"ur Astrophysik, Karl-Schwarzschild-Strasse 1, 85748 Garching, Germany}
\author{A. Eff-Darwich}
\affil{Departamento de Edafolog\'ia y Geolog\'ia, Universidad de La Laguna, La Laguna, Tenerife, Spain}
\author{R.A. Garc\'ia}
\affil{Laboratoire AIM, CEA/DSM-CNRS-Universit\'e Paris Diderot; CEA, IRFU, SAp, Centre de Saclay, F-91191, Gif-sur-Yvette, France}
\author{S.J. Jim\'enez-Reyes}
\affil{Instituto de Astrof\'isica de Canarias, 38205, La Laguna, Tenerife, Spain}
\author{S.G. Korzennik}
\affil{Harvard-Smithsonian Center for Astrophysics, Cambridge, MA 02138, USA}
\author{S. Turck-Chi\`eze}
\affil{Laboratoire AIM, CEA/DSM-CNRS-Universit\'e Paris Diderot; CEA, IRFU, SAp, Centre de Saclay, F-91191, Gif-sur-Yvette, France}

\begin{abstract} 
The detection of the signature of dipole gravity modes has opened the path to study the solar inner radiative zone. Indeed, $g$ modes should be the best probes to infer the properties of the solar nuclear core that represents more than half of the total mass of the Sun.
Concerning the dynamics of the solar core, we can study how future observations of individual $g$ modes could enhance our knowledge  of the rotation profile of the deep radiative zone. Applying inversions on a set of real $p$-mode splittings coupled with either one or several $g$ modes, we have checked the improvement of the inferred rotation profile when different error bars are considered for the $g$ modes.
Moreover, using a new methodology based on the analysis of the almost constant separation of the dipole gravity modes, we can introduce new constraints on solar models. For that purpose, we can compare  $g$-mode predictions computed from several models including different physical inputs with the $g$-mode asymptotic signature detected in GOLF data and calculate the correlation. This work shows the great consistency between the signature of dipole gravity modes and our knowledge of p-modes:  incompatibility of data with a present standard model including the Asplund composition.
\end{abstract}


\section{Introduction}

\begin{figure}[htbp]
\begin{center}
\includegraphics[width=8cm]{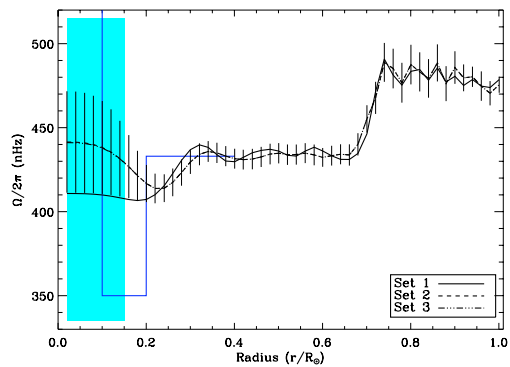}
\caption{Results of the inversion of a simulated rotation profile (blue) using only $p$ modes with three different sets of modes (see Garc\'ia et al. 2008b).}
\label{pinversion}
\end{center}
\end{figure}

Solar modeling still needs improvements as constraints are not strong enough to correctly reconstruct for instance, the observed sound speed profile as well as the rotation profile in the solar core. \\
With the release of the new abundances of Asplund et al. (2005), the discrepancy between the sound speed and density profiles of standard models including this new composition and the ones inferred from acoustic ($p$)-mode observations increased, especially in the radiative zone (Turck-Chi\`eze et al. 2004a; Bahcall et al. 2005; Guzik et al. 2005; Mathur et al. 2007; Zaatri et al. 2007). Many attempts have been made to explain this discrepancy by attributing it to opacities, equation of states, or by changing the composition in Ne or the microscopic diffusion rates (see Basu et al. 2008; Guzik et al. 2008 in these proceedings).\\
As for the rotation profile, it is well established down to 0.3R$_\odot$ (Couvidat et al. 2003; Garc\'ia et al. 2007) with larger and larger uncertainties if we go deeper into the solar core. Besides, the limitation of the $p$ modes to bring information on the solar core rotation has been shown with inversions of data sets using real error bars (Garc\'ia et al. 2008b). Indeed, by using more low-degree $p$ modes (up to 3.4 $\mu$Hz), we manage to reconstruct artificial rotational profiles down to 0.2$R_\odot$ (see Fig.~\ref{pinversion}).

\noindent Gravity ($g$) modes would be the best data to improve our knowledge on the dynamics and the structure of the solar core. Today, we are almost 'touching' their individual detection and with a few more inches, we would be able to grab them. Here is a non exhaustive list of works dedicated to the search for $g$ modes: detection of one candidate at 220 $\mu$Hz (Garc\'ia et al. 2001; Gabriel et al. 2002, Turck-Chi\`eze et al. 2004b; Mathur et al. 2007), the component of this candidate followed in time with VIRGO (Garc\'ia et al. 2008; Jimen\'ez \& Garc\'ia 2008 these proceedings), the detection of their asymptotic properties (Garc\'ia et al. 2007). Moreover, new techniques are still being developed (Appourchaux et al. 2008; Garc\'ia et al. 2008; Grec et al. 2008 these proceedings). In this work, we investigated what can be inferred about the solar core from $g$-mode detection.

\section{On the Dynamics of the Inner Radiative Zone}

\begin{figure}[htbp]
\begin{center}
\includegraphics[width=8cm]{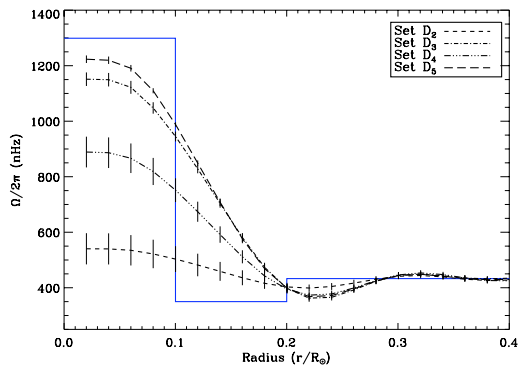}
\caption{Inversions for the simulated profile (blue). In the four sets of data, $p$ modes used are as set 1 of Fig.~\ref{pinversion} (see Mathur et al. 2008).} 
\label{ginversion}
\end{center}
\end{figure}

To estimate what specific information on the core rotation is added by $g$ modes, modes for which we have today candidates and that are likely to be unequivocally observed in the future, we used here an inversion code based on the Regularized Least Square methodology (Eff-Darwich \& P\'erez-Hernandez 1997). We created an artificial rotational profile, which is  a step-like profile (see Fig.\ref{ginversion}). From this profile, we calculated artificial splittings and we inverted them after attributing them observational error bars. We have considered two sets of  $g$ modes: one with the current $g$-mode candidate (D$_2$ and D$_3$) and another one with 8 low-degree $g$ modes ($\ell$ = 1, $n$ = -2 to -5 and $\ell$ = 2, $n$ = -3 to -6) (D$_4$ and D$_5$), and for both cases uncertainties values of 75 (D$_2$ and D$_4$) and 7.5 nHz (D$_3$ and D$_5$).

\noindent Fig.~\ref{ginversion} shows that with one $g$ mode and a very large error bar,  we do not have any information on the behavior of the rotation in the solar core. Then, either by increasing the number of modes or by decreasing the error bars, we start to have some indication of the actual simulated rotation rate in the core. However, such small amount of modes is not enough to reconstruct the profile. We only have one or two points below 0.1R$_\odot$ in the inversion. We need more $g$ modes to constrain the profile at these depths. For the region between 0.1 and 0.2 R$_\odot$, low-degree high-frequency $p$ modes are needed as their inner turning points are in this region. 

\begin{figure}[htbp]
\begin{center}
\includegraphics[width=8cm]{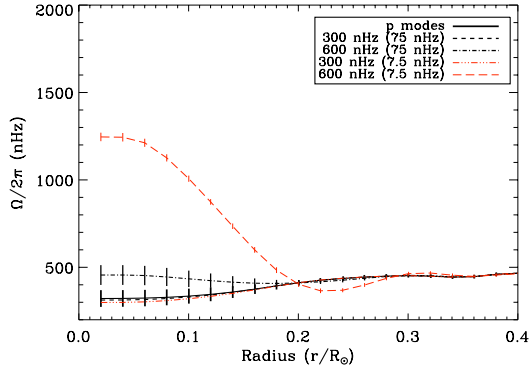}
\caption{Inversions of real data with only $p$ modes and adding the $g$-mode candidate (from Mathur et al. 2008).}
\label{default}
\end{center}
\end{figure}

\noindent Then, we applied the inversion to real data using only $p$ modes and also adding the $g$-mode candidate with the previous values of uncertainty. As different scenarios were proposed to explain this candidate, two values of splitting have been used: 300 and 600 nHz. Whatever the value of the splittings is, the result is stable and we have the same behavior down to 0.2R$_\odot$. If we take into account the error bars, the rotation is well defined down to 0.17R$_\odot$. However, the rotation profile below this radius cannot be determined as long as we do not detect several $g$ modes individually.

\section{On the Physical Inputs in Solar Models}

In this section, we tried to see how the detection of asymptotic properties could provide us some information on a few physical inputs in solar models. We know that $g$-mode periods are periodically spaced in the Power Spectrum Density (PSD). The Fourier transform of this PSD (PS2) has a large peak ($\Delta P_1$ for the modes $\ell$=1) corresponding to this periodicity (see for example Garc\'ia et al. 2008c). Using the same methodology as the one developed in Garc\'ia et al. (2007), we applied it to seven solar models with different physical inputs. Depending on the models used, $g$-mode frequencies can vary up to 5$\mu$Hz. This information cannot be used as long as $g$ modes are not detected individually. But the asymptotic value of $\Delta P_1$ is within a one-minute range depending on the physical input of the model (Mathur et al. 2007). 

\begin{figure}[htbp]
\begin{center}

\begin{tabular}{cc}
\includegraphics[height=5.5cm,width=6cm]{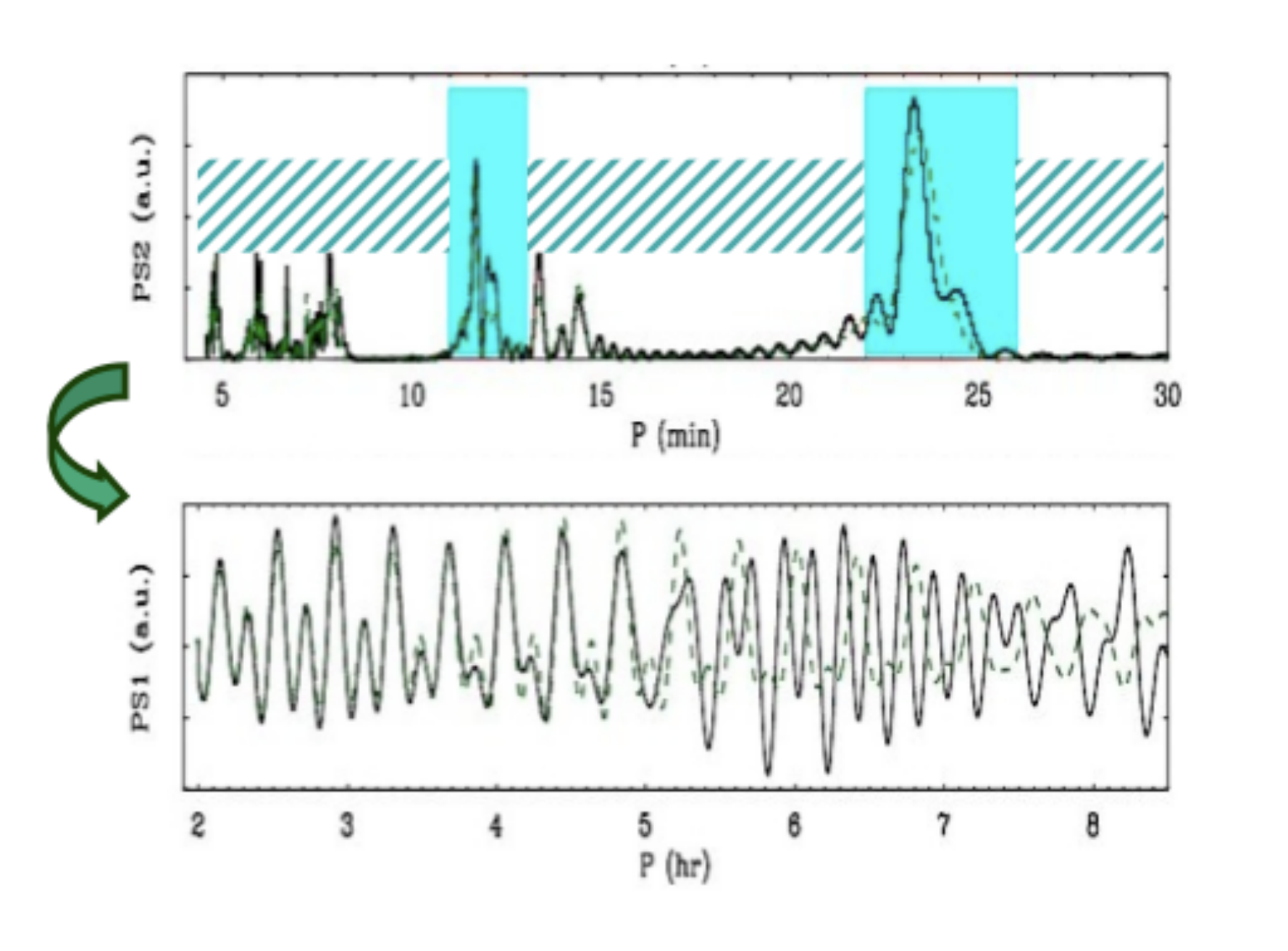}&
\includegraphics[height=5.2cm,width=6cm]{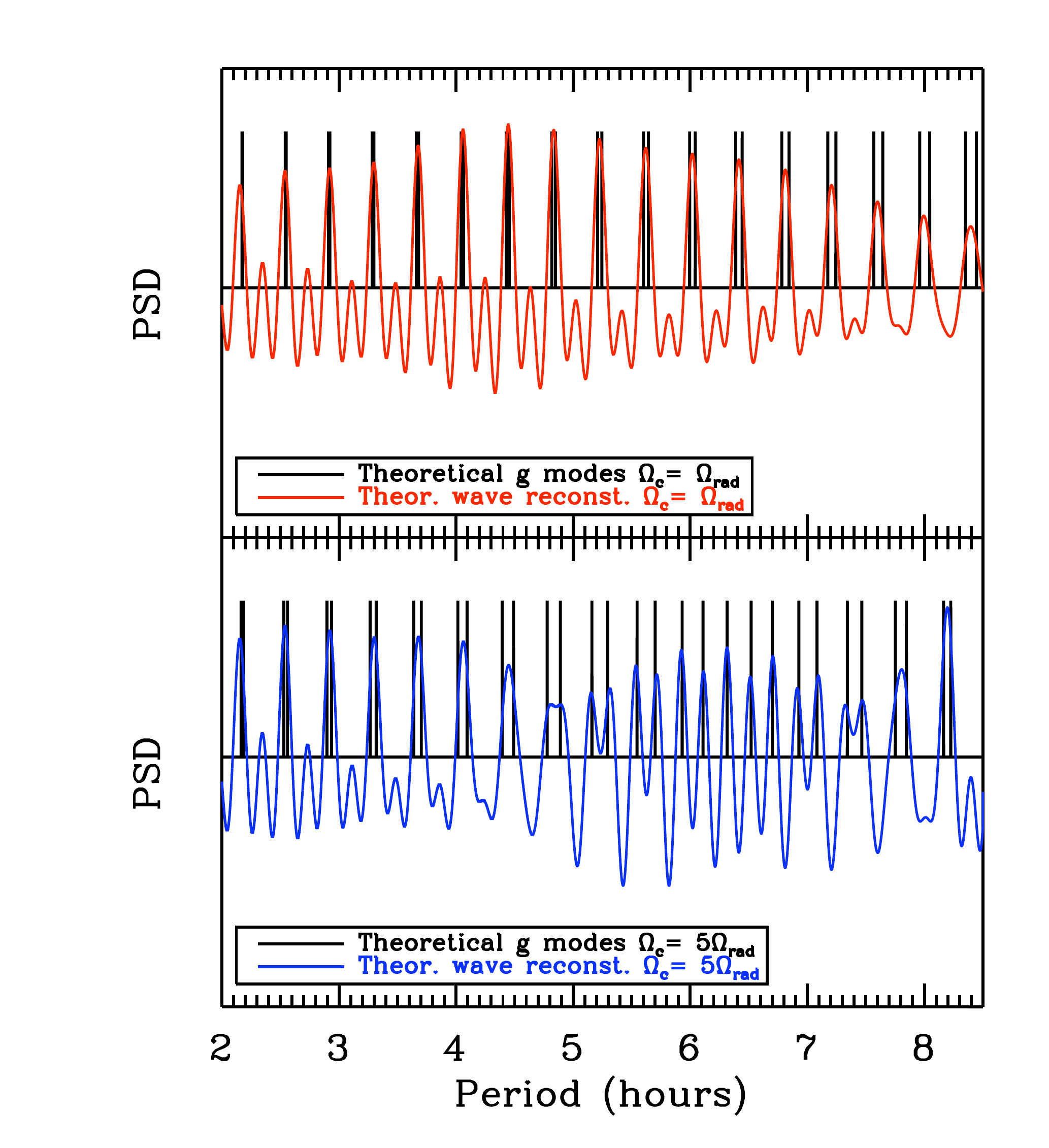}
\end{tabular}

\caption{Reconstruction of signal: on the left (top), PS2 for 2 models with different rotation profile, as a function of period that is filtered on the peak of $\Delta P_1$ and its first harmonic (blue regions). On the left (bottom), signal reconstructed for both models. On the right, simulated signal (black peaks) and signal reconstructed for 2 rotation profiles (red and blue) (extracted from Garc\'ia et al. 2007).}
\label{recons}
\end{center}
\end{figure}

\noindent The method consists in filtering only the region delimited by the peaks of the $\Delta P_1$ and its first harmonic in the PS2 (Fig. \ref{recons} left) so that we can reconstruct the signal corresponding to this region of the spectrum. This method is very powerful as it reconstructs the components of the modes (Fig. \ref{recons} right) and it is sensitive to the rotation. Indeed, in the PS2, we cannot distinguish between a model with a solid rotation and a model with a core rotating faster than the rest of the radiative zone. Thus, as well as testing the dynamics of the core, we can check the sensitivity to some physical processes. Two parameters are used: the physical input such as the chemical composition, the microscopic diffusion or the diffusion in the tachocline, and the rotation profile which is a step-like profile with different values of rotation rate in the core. 

\begin{figure}[htbp]
\begin{center}
\includegraphics[width=8cm]{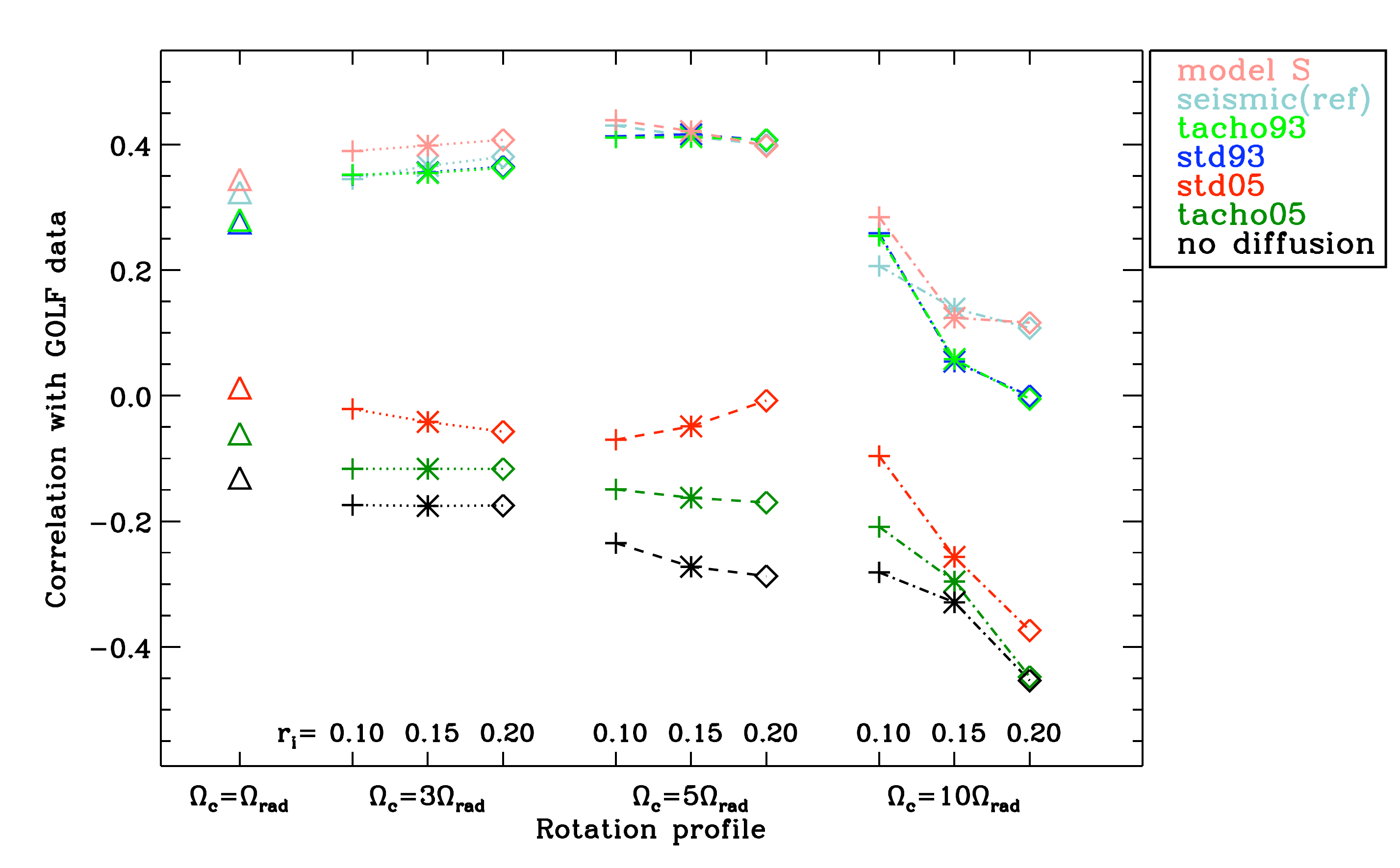}
\caption{Correlation rates between reconstructed signal of GOLF data and seven solar models as a function of the rotation profile (from Garc\'ia et al. 2008c).}
\label{corr_rate}
\end{center}
\end{figure}

\noindent Once we have reconstructed the signal corresponding to different solar models and to the GOLF data, we calculate the correlation rate between these models and the data. Fig. \ref{corr_rate} shows the correlation rates for the seven models as a function of the rotation profile. For a given rotation profile, two groups can be distinguished: the first one with higher correlation rates for the models using the old abundances of Grevesse et al. (1993) and another group of models with the abundances of Asplund et al. (2005) with very low correlation rates (even negative) and the model without microscopics diffusion. Thus, GOLF data are more compatible with models S, and with the old abundances than the other models. Then, a rotation profile with a high rate from 3 to 5 times the rate in the rest of the radiative zone gives higher correlation rates.

\section{Conclusions}

On one hand, a theoretical work was carried to see how individual $g$-mode detection could improve our knowledge on the rotation of the solar core. We showed that with one or a few $g$ modes, we can have some information on the rotation rate in the core. But to better reconstruct the rotation profile in the core, we need more modes.
On the other hand, we used the detection of asymptotic properties to extract information on the structure of solar models. The methodology used is sensitive to the structure for a few models and also to the rotation profile. Moreover, the results obtained are in agreement with $p$-mode results as GOLF data are more compatible with standard models using old abundances. We still have to keep in mind that these standard models are not complete and the discrepancy is not necessarily related to the abundances themselves.
However, to improve our knowledge on the solar burning core, we need to detect these $g$ modes: individually, we would be able to constrain the dynamics and globally, we would have some constraints on physical inputs.

\acknowledgements 
This work has been partially funded by the Spanish grant PENAyA2007-62650 and the CNES/GOLF grant at the SAp-CEA/Saclay. 



\begin{thebibliography}{}

\bibitem[]{} Asplund, M., Grevesse, N. \& Sauval, A.~J. 2005, Cosmic Abundances as Records of Stellar Evolution and Nucleosynthesis, 336, 25

\bibitem[]{} Bahcall J.N., Serenelli, A.M. \& Basu, S. 2005, ApJ Lett., 621, 85

\bibitem[]{} Couvidat, S., Garc\'ia, R.A., Turck-Chi\`eze, S., et al.  2003, Sol.Phys., 597, 77

 \bibitem[]{} Eff-Darwich, A. \& P\'erez-Hernandez, F. 1997, A\&A, 125, 1

\bibitem[]{} Gabriel, A.H., Baudin, F., Boumier, P., et al. 2002, A\&A, 390, 1119

\bibitem[]{} {Garc{\'{\i}}a}, R.~A., {R\'egulo}, C., {Turck-Chi\`eze}, S., {et~al.} 2001,
  \solphys, 200, 361

\bibitem[{{Garc{\'{\i}}a} {et~al.}(2007){Garc{\'{\i}}a}, {Turck-Chi\`eze},
  {Jim\'enez-Reyes}, {Ballot}, {Pall\'e}, {Eff-Darwich}, {Mathur}, \&
  {Provost}}]{2007Sci...316.1591G}
{Garc{\'{\i}}a}, R.~A., {Turck-Chi\`eze}, S., {Jim\'enez-Reyes}, S.~J.,
  {et~al.} 2007, Science, 316, 1591
  
  \bibitem[{{Garc{\'{\i}}a} {et~al.}(2008{\natexlab{a}}){Garc{\'{\i}}a},
  {Jim{\'e}nez}, {Mathur}, {Ballot}, {Eff-Darwich}, {Jim{\'e}nez-Reyes},
  {Pall{\'e}}, {Provost}, \& {Turck-Chi{\`e}ze}}]{2008AN....329..476G}
{Garc{\'{\i}}a}, R.~A., {Jim{\'e}nez}, A., {Mathur}, S., {et~al.}
  2008{\natexlab{a}}, Astronom. Nachrichten, 329, 476
  
  
  \bibitem{}
{Garc{\'{\i}}a}, R.~A., Mathur, S., Ballot, J., et al. 2008b,
  \solphys, 251, 119
  
   \bibitem[]{}
{Garc{\'{\i}}a}, R.~A., {Mathur}, S. \& Ballot, J.
  2008c, \solphys, 251, 135
  
  \bibitem[]{} Grevesse \& Noels 1993, Origin and Evolution of the Elements, p15
  
  \bibitem[]{} Guzik, J.A., Watson, L.S. \& Cox, A.N. 2005, ApJ, 627, 1049
  
  \bibitem{} Mathur, S., Turck-Chi\`eze, S., Couvidat, S. \& Garc\'\i a, R.A. 2007, ApJ, 668, 594 

\bibitem[]{} Mathur, S., Eff-Darwich, A., Garc\'ia, R.A. \& Turck-Chi\`eze, S., 2008, A\&A, 484, 517

\bibitem[]{} Turck-Chi\`eze, S., Couvidat, S.,  Piau, L., et al., 2004a, Phys. Rev. Lett., 93, 1102

\bibitem[]{} Turck-Chi\`eze, S., Garc\'ia, R.A., Couvidat, S.,  et al., 2004b, ApJ, 604, 455

\bibitem[]{} Zaatri, A., Provost, J., Bertomieu, G., et al., 2007, A\&A, 265, 115

%
%
%
%
%
%
%
\end{thebibliography}
\end{document}